# A short review of the main concerns in A.I. development and application within the public sector supported by NLP and TM[*]

**Carlos Ferreira**[**]
*Faculty of Engineering*
*University of Porto*
ferreira.carlos@fe.up.pt

**Abstract:** *Artificial Intelligence is not a new subject, and business, industry and public sectors have used it in different ways and contexts and considering multiple concerns. This work reviewed research papers published in ACM Digital Library and IEEE Xplore conference proceedings in the last two years supported by fundamental concepts of Natural Language Processing (NLP) and Text Mining (TM). The objective was to capture insights regarding data privacy, ethics, interpretability, explainability, trustworthiness, and fairness in the public sector. The methodology has saved analysis time and could retrieve papers containing relevant information. The results showed that fairness was the most frequent concern. The least prominent topic was data privacy (although embedded in most articles), while the most prominent was trustworthiness. Finally, gathering helpful insights about those concerns regarding A.I. applications in the public sector was also possible.*

**Keywords:** *artificial intelligence, public sector, data privacy, ethics, interpretability, explainability, trustworthiness, fairness*

## 1 INTRODUCTION

Artificial Intelligence (A.I.) is not a new subject. Indeed, A.I. started in the 1950s [1]. The field has passed for different evolutions, such as the development of the Stochastic Estimation (1952) [2], the Perceptron (1968) [3], and Backpropagation – Multi-Layer Perceptron (1986) [4]. Nevertheless, only in the last two decades, A.I. could be disseminated through different sectors of society, such as business, industry and public sector, mainly due to three factors: First, the exponential development of tools to collect and store immense volumes and variety of data, also known as BigData [5]. Second, the development of powerful Graphics Processing Units (GPUs) [6], [7]. Third, the development and release of open-source packages such as TensorFlow and PyTorch [8].

Businesses have applied A.I. aiming to simplify their operations, enhance predictions and insights, and make decision-making intelligent, generating better results, not only in terms of revenue. These objectives can be achieved through A.I. applications by simulating and predicting human behaviour. For instance, A.I. models can predict customer preferences, navigation patterns, grouped interests, and purchasing behaviour. On the other hand, these predictions are used to improve business decision-making [9].

In industry, A.I., jointly with other approaches such as Cyber-Physical Systems (CPS), Internet of Things (IoT), Internet of Services (IoS), and Data Analytics, has favoured changes in the maintenance paradigm (from reactive to proactive maintenance actions), besides leveraged and more easily applications of approaches such as Zero-Defect Manufacturing (ZDM) and Condition-Based Maintenance (CBM). ZDM ensures process and product quality by reducing defects, leading to an improvement in manufacturing sustainability. CBM allows scheduling maintenance routines in advance, generating more effective planning, improving safety by preventing performance degradation and failures, creating higher operation reliability, increasing production, and minimising costs due to reduced downtime [10].

In the public sector, A.I. has been used in different ways. In 2020, Missuraca & van Noordt released the report "Overview of the use and impact of A.I. in public services in the E.U." as a part of the "A.I. Watch – Artificial Intelligence in public services". The report describes and analyses 230 cases of A.I. applications in the public sector. These cases include but are not limited to planning and paying social benefits, immigration control, fraud detection, infrastructure projects planning, citizen support (answering queries), court issues, hospital triage, and smart city services [11].

---

[*] **Preprint:** Under review
[**] Ph.D. student in Engineering and Industrial Management and Research Group Member at Center for AI and Digital Policy (CAIDP) since January 2023.

However, all these domains of A.I. applications have some concerns. Some examples that have been more intensively observed in the last few years are data privacy, ethics, interpretability, explainability, trustworthiness, and fairness [12]–[14]. This article has two fundamental objectives. First, to test the use of basic concepts of Text Mining (TM) and Natural Language Processing (NLP), two techniques embedded in the universe of A.I., to help retrieve potential research papers published in conference proceedings in ACM Digital Library and IEEE Xplore in the last two years. Then, based on the results, it aims to briefly review these concerns in the public sector context and provide helpful insights.

The remaining work presents the methodology in Section 2 and the results in Section 3. Next, Section 4 discusses some findings. Finally, Section 5 concludes this work and gives some future research directions. In the Appendix, the readers can find the codes used in the search.

## 2 METHODOLOGY

The methodology adopted to achieve the objectives of this work contains two parts. First, an advanced search (i.e., using a search query containing Boolean Operators and Wildcards) was performed, and filters were applied. Second, basic concepts of NLP and TM were applied to help the final selection of the papers to analyse thoroughly.

### 2.1 Advanced search

The advanced search was performed in the Database of the Association for Computing Machinery Digital Library (ACM Digital Library) and IEEE Xplore. These databases were chosen since it converges more than 200 conferences, symposiums and workshops yearly, with many of these events focused on the public sector. The search query used was ("artificial intelligence" OR "A.I.") AND (ethic* OR fair* OR "data privacy" OR interpreta* OR expla* OR trust*) AND ("public sector"), and after retrieving the results, three filters available in the website were applied as follows.

First, only the results published in the last two years were selected. This timeframe is reasonable since advances in this area have occurred at a breakneck pace. Therefore, information published over two years is highly likely obsolete. For instance, according to [13], in October 2017, only Finland published its A.I. National Strategy among the EU27+NO. In January 2020, there were already 15 A.I. National Strategies published. Finally, in January 2022, 24 A.I. National Strategies were published, four being updated and four in progress.

Second, only research articles (filter) published in the proceeding conferences (filter) were selected. Conferences are crucial opportunities for presenting and discussing research in different phases (conception, research, implementation, and updating). The results of these presentations and discussions are then published as proceedings, which can effectively capture cutting-edge innovation. In this context, the proceedings from ACM Digital Library and IEEE Xplore can competently capture innovations and concerns in A.I. applied to the public sector.

The search query returned 587 results. Filtering the last two years returned 263 results. Filtering by proceedings returned 216, and filtering by research articles returned 163 results. These results were still refined as described next. Table 1 presents the top five conferences.

Table 1: Top five conferences retrieved

| Proceedings/Book Names | Papers |
|---|---|
| **ICEGOV**: International Conference on Theory and Practice of Electronic Governance | 29 |
| **DG.O**: Annual International Conference on Digital Government Research | 21 |
| **CHI**: Conference on Human Factors in Computing Systems | 15 |
| **FAccT**: ACM Conference on Fairness, Accountability, and Transparency | 14 |
| **AIES**: AAAI/ACM Conference on A.I., Ethics, and Society | 11 |

### 2.2 NLP and TM to help selection

Once the advanced search results were retrieved, it was possible to implement some basic concepts of TM (search for words and expressions in the body of the text) and NLP (word count and metric calculation) to help in the final selection of papers. These concepts were implemented by using IDLE Python 3.10.5, with libraries Pandas (Version: 1.5.3), Fitz (Version: 0.0.1.dev2), RegEx (Version: 2022.6.2) and O.S. Besides, the code was implemented using Intel(R) Core(T.M.) i5-10210U CPU @ 1.60GHz   2.11 GHz, 16.0 GB (15.8 GB usable), 64-bit operating system, x64-based processor, Windows 10 Pro, Version 22H2. The implementation occurred through Code 1 (search_file.py), Code 2 (count_words_ACM_searched_filtered.py), and Code 3 (count_words_ACM_searched_selection.py), which are presented in Appendices A.1, A.2, and A.3. In the codes, libraries are highlighted in orange, comments in red (and marked with the symbol # at the beginning) and lines of code themselves, were kept in black.

First, the 163 papers were treated twofold: (i) sections "Acknowledgements" and "References" were removed from the documents. These sections do not provide any relevant content. Besides, the search words are commonly repeated in the title of the papers in the references, which could integrate some bias to the results, and (ii) if the title of the article contained some search word and it was in the header or footer of the paper, these elements were removed, to avoid double counting of the words.

Second, searching for likely and relevant words and expressions in the body text of the 163 selected articles was performed. The group of search expressions consisted of {"artificial intelligence" and "public sector"}. The group of search words consisted of {"ethics", "ethical", "fair", "fairness", "data privacy", "interpretable", "interpretability", "explainable", "explainability", "trust", "trustworthy", "trustworthiness"," responsible A.I.", and "transparency"}. Code 1 performed the search so that, to select a paper, it had to contain both the search expressions and at least one of the search words, no matter which one. The result was a set with 85 selected papers.

Third, the number of occurrences of the search words within the selected papers was verified in the following way. Code 2 counted the number of occurrences of each search word in the body of all texts. Then, it calculated the average of occurrences without considering the papers where the search word did not occur. The average was also rounded to zero digits. Finally, for each search word, it returned only the articles where its occurrence was greater than or equal to its average throughout the papers. At the end of this process, Code 2 returned a list with several articles overlapped because the same word had a high count on different papers, making the article selected more than once. To correct the overlapping, the M.S. Excel function was used to remove duplicates based on the paper's name. The result was a set with 44 included papers to analyse thoroughly. Figure 1 shows a schematic workflow for the selection and inclusion process.

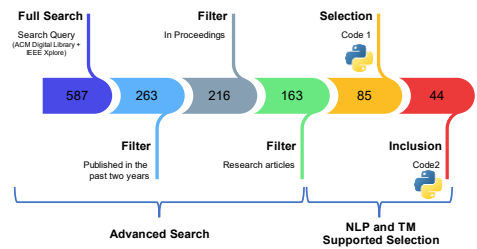

*Figure 1: Schematic workflow of the selection and inclusion process*

## 3 RESULTS

The first result regards the most relevant concern throughout the included papers. Still using computational resources and NLP concepts, the search words were counted in the body of the selected documents using Code 3. Figure 2 shows a cloud of words that allows perceiving the relevance of each search word, especially highlighting the concern with fairness and ethics.

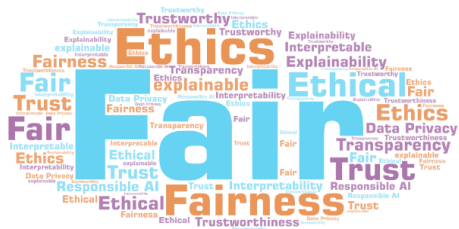

*Figure 2: Cloud of words for the set of search words considered in this work.*

The second result regards the main topic in the included papers. The papers were evaluated and categorised according to the most prominent subject in the body of the text. The most occurred word in the body of the text was considered to determine the topic. Also, due to the similarity among the search words, the following interpretation was regarded: ethical and ethics, subject ethics; explainable and explainability, subject explainability; fair and fairness, subject fairness; interpretability and interpretable, subject interpretability; trust, trustworthiness and trustworthy, subject trustworthiness. Figure 3 shows this result graphically.

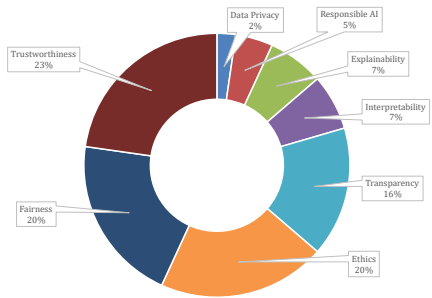

*Figure 3: Most prominent subjects in the included papers.*

Once subsets based on the subjects were constructed, it was possible to analyse the papers more in-depth. The analysis was performed without using computational resources, even though it is possible. The third result comprises the most important insights about each subject, presented in the following sections, sorted in ascending order according to the share of topic appearance.

### 3.1 Data Privacy

Digital transformation in public institutions can assume different parameters and stages. One of these parameters is information security and data privacy. In the first stage (Silo System Stge), public institutions are generally incapable and do not have tools for managing data privacy issues. In the second stage (Engaged Stage), they are still constructing the necessary capabilities or have developed and applied a few mechanisms to deal with data privacy issues and compliance. In the third stage (Optimised Stage), they have awareness and structured capacity (tools and infrastructure) to manage data privacy issues. Finally, in the last stage (Transformation Stage), once all digital services (including A.I. applications) are designed and implemented, data privacy becomes critical in achieving a highly matured level in public institutions [15].

### 3.2 Responsible A.I.

Humans develop the technology (e.g., programmers, computer scientists, and data analysts) and can share the responsibility for potential damages to their developed systems. However, they should not do the work of other experts and specialists. They must work together as a network to develop policies and practices to anticipate and reduce risks. This network, which includes national and state legislatures, government agencies, professional societies and the private and public sectors, should build a culture of responsible A.I. development [16].

Public organisations developing ethical guidelines and legal/regulatory rules for A.I. systems can be considered to play an indirect role in the behaviour of responsible A.I. systems. Further, although this development has presented a consensus that the legal obligations should be with humans, with A.I. systems' increasing sentience and autonomy, regulations and public policies on legal obligations of A.I. responsibilities should adapt beyond human ethics [17].

### 3.3 Explainability

The total investments in A.I. and these technologies' revenues have increased over five times from 2017 to 2021. The public sector (as in other domains of society) has reached high-level A.I. tools, some of them bringing undesirable effects for citizens (and public organisations), and must justify its decisions, primarily when these decisions target citizens' safety and rights. The explainability deals with this concern, and its absence leads to low accountability, fairness, responsibility and transparency in AI-based decision-making [18]. Nevertheless, the ML model governance is challenging due to the own nature of these models, which leads to a lack of explainability. In this way, although these models need to be (and usually to be) tested through numerous ways to ensure their operability, their transparency and explainability should be increased [19].

In the backstage of these decisions, deep learning models usually guide them, with these models, in different domains, falling into the "black box" ("high level of difficulty for the system to provide a suitable explanation for how it arrived at an answer"), which should be deciphered (on different levels) to ensure accountability and transparency. In this context, XAI tools have become essential approaches to overcoming this problem by increasing the model's explainability and making deep learning models easily understandable to humans and more comprehensively governed by public and societal involved actors [18], [19].

### 3.4 Transparency

Information and Communication Technologies (ICT) governance has practices and structures of responsibility, which are required for public policies and government services management. These elements can enhance transparency, decision-making, and accountability and derive from standards and frameworks such as Information Technology Infrastructure Library (ITIL), Control Objectives for Information and Related

Technologies (COBIT) and International Organization for Standardization (ISO) 38500 [20].

The public sector should not escape these elements, as it must justify its decisions and keep its decision-making processes visible, explainable, auditable, accountable and transparent. On the other hand, citizens have the right (and obligation) to verify whether public decisions affecting them have been taken inside or beyond the limits of the law and trigger the justice apparatus if necessary. In this context, and considering the significant challenges in implementing these values (due to the nature of A.I. algorithms) and discretion still being an essential element in decision-making, some social groups have faced the problem, and some solutions (legal, technological and innovative) have been put on the table [21].

Sequential Decision-Making (SDM) algorithms have been proposed having the potential to enhance transparency (identify dangerous feedback loops and biases in existing systems), efficiency (identifying bottlenecks in the delivery of essential services, potential corruption and fraudulent outflows) and adaptability of Government (helping to narrow the divide between the ex-post and ex-ante camps) [22]. Also, cities such as Geneva, Amsterdam and Helsinki have implemented platforms for registering algorithms, improving the transparency about using A.I. algorithms in the cities [23]. Further, although some limitations exist, A.I. audits considering third parties have achieved robust oversight of A.I. systems and improved transparency [24], [25]. Additionally, frameworks have been introduced, offering a systematic approach for enhancing transparency and accountability of Automated and Algorithmic Decision-Making (ADM) through Machine Learning (ML) [26].

Finally, besides the social groups and tools, another aspect that can be highlighted in times of Big Data and AI-augmented public sector regards data collection, storage and transformation. Open data environments are essential in the public sector to ensure transparency and public value using new technologies, so they should be enhanced [27]. Still within that context, for example, in healthcare and public welfare areas, the role of the human(s) behind the data, practices of data scientists, ML developers, and data annotators, and the behaviour of the data from their collection to their application in ML models when multiples persons manipulate different versions of the same datasets have been studied aiming to improve the transparency of A.I. systems applied to them [28].

### 3.5 Ethics

Ethical behaviour "requires consideration of how to treat others, and what becomes of others and oneself in addressing intermediate problems" [29]. This issue has been the most prevalent core pillar in developing A.I. policies and the second area requiring more research [30]. For instance, the ethical implications of Big Data (due to individual rights) are the core challenge in the biomedical domain, while it is, at the same time, the critical driver of research and development in that field [31]. Another area of high potential for ethical implications (due to the possibility of human rights violations) is the judiciary [32].

From the point of view of organisations, due to the explosive growth of A.I. applications, international organisations such as OECD and the Council of Europe and agencies such as UNESCO have developed broad documents containing embedded ethical issues [33]. In addition, different public and private institutions, besides non-governmental organisations (NGOs), have also developed documents facing ethical concerns regarding the development of A.I. in their domains.

From the point of view of human resources, developers (researchers and practitioners) who work with technology design have been increasingly required to be ethically engaged for several reasons, such as the legal and societal implications of technological innovation and power-knowledge asymmetries. In this context, professional organisations also have launched ethical codes such as the ACM Code of Professional Conduct and IEEE Code of Ethics linked to technology development [29].

These instruments, as well as guidelines and principles, have become essential pieces of the ethical implementation process, although they are more focused on the fair and trust use and implementation of A.I. systems, being more generic ethically [34], [35]. Nevertheless, other factors, such as lack of equal opportunities and income and wealth inequality, should be approached, strongly emphasising ethics and responsible A.I. [30], while the role of the public sector as a critical stakeholder should be improved [32]. Another proof of this mandatory improvement of the public sector role relies upon their documents being generated through more participatory processes and being more engaged with law and regulation when compared with the private sector documents. This balance (participation and regulation) can smooth out specific conflicts, such as experts versus public representatives and social versus economic interests [36].

Since ethics is an umbrella and covers different concerns such as trust, accountability, fairness and bias, it should not be approached individually but, instead, jointly with other considerations (such as those highlighted in this work) to improve A.I. public systems and ensure human-centricity and public benefits with progressiveness and sustainability [29], [37], [38]. Finally, tools and processes such as those described in [39] can be helpful to ensure that ethical aspects are being incorporated into A.I. systems and evaluate the trustworthiness of these systems.

## 3.6 Fairness

The lack of algorithmic fairness can be considered an issue derivated from a misunderstanding about A.I. being part of a comprehensive paradigm involving technical and social discussions. This misunderstanding can cause two types of situations: the framing trap – a "failure to model the entire system over which a social criterion, such as fairness, will be enforced" and the formalism trap – a "failure to account for the full meaning of social concepts such as fairness, which can be procedural, contextual, and contestable, and cannot be resolved through mathematical formalisms" [40].

The literature has widely pointed out how adopting unfair algorithms can lead to disparate impact and other negative consequences [41]. However, despite the existence of increasing research on algorithms that can guarantee fairness, they have not been widespread in A.I. systems throughout different areas (including the public sector), mainly due to the trade-off between fairness and predictive performance since fairness standards and ML loss functions influence learning behaviour competitively [42], [43].

To mitigate the problem and enforce algorithmic fairness, some countries such as The United Kingdon, Germany, China and The United States and organisations such as the Institute of Electrical and Electronics Engineers (IEEE) and the International Organization for Standardization (ISO) have launched laws and standards to be followed internally and by associated professionals. Also, the European A.I. Act has been intensively discussed to regulate all artificial intelligence (A.I.) systems used in the European Union by private and public organisations [44].

Finally, in this context, some approaches have been developed to deal with and improve fairness in A.I. applications (extensively to the public sector), such as Vertical Equity (appropriately considering relevant differences across individuals) [41], the Full Hyper-parameter Search (FHS) method (for selecting the best fair models regarding performance and the best fairness metric) [42], and the incorporation of a Stakeholder-centred Design Approach for A.I. in Human Resource Management (HRM) [45]. Further, some prominent "open source fairness toolkits" such as sci-kit-fairness/sci-kit-lego, IBM Fairness 360 (AIF360), Aequitas Tool, Google What-if tool, and Fair-learn are available aiming to assess and mitigate unfair outcomes in algorithms, despite the existence of gaps between their capabilities and the real needs of practitioners [46].

## 3.7 Trustworthiness

The public service is facing an enormous and fast change in its policy-making, governance and delivery mechanisms due to the increasing digitalization (sometimes called e-Government or E-governing), allowing all stakeholders to access more efficient, effective and user-centric systems [47], [48]. When powered by A.I. (mostly noted by machine learning, sentiment analysis, text mining, and pattern recognition), it can potentially be disruptive in several aspects, but especially in the interaction between the citizens and the Government [49].

Indeed, this digitalization is already impacting the improvement of administration, public services and social value in several ways due to the recognized embedded trustworthiness in these systems, mainly by the citizens [50]. However, the potential sources of lack of trustworthiness and other attributes in these systems need increasing focus on research since they can have a relevant impact, for instance, on political definitions through participation channels such as voting [47]. In this context, another area that can suffer from a lack of trustworthiness in the context of A.I. application is the criminal justice system. It is a sensitive area since untrust algorithms can affect fundamental rights like freedom. For this reason, principles such as fit for purpose, data quality, fairness, transparency and accountability, privacy, robustness, and contestability should be incorporated to ensure the trustworthiness of these algorithmic systems [51].

Besides all the previous considerations, and although the current A.I. system's benefits and potential future ones, some challenges should be overcome to ensure the necessary trust of society. First, there is a need to increase technical human resources capable of developing and deploying A.I. systems. Then, public decision-makers should remember that there is a gap in the societal digital culture (that should be reviewed), which, added to an excess of confidence in A.I. predictions, might be catastrophic (e.g., in health care systems). Finally, the more systems are implemented, the more resources (technical, human and financial) are needed to maintain them. The public authorities must consider this, especially when allocating the required financial resources to implement these A.I. systems effectively [52].

Finally, it is essential to highlight the success of digitalization (that includes A.I. systems and other technologies) implementation in the public sector is strongly dependent on factors such as economic and social conditions, legal foundations, technical infrastructures and perceived public trust, which should be achieved gradually throughout the process of deployment of these digital systems [53].

## 4 DISCUSSION

### 4.1 Limitations

The limitations of this work rely mainly upon two points. First, the NLP analysis that supported the selection considered only the word but not its context within the text. It was a recognised weakness that needs to be mitigated in future work. Second, the research was performed in only two databases; hence, the number of selected articles was small, which may be insufficient for generalising the results captured in this work more robustly.

### 4.2 Methodology

Due to the NLP analysis weakness detected and previously pointed out, a complimentary analysis was performed twofold, aiming to avoid the introduction of biased results and remarks in this work.

First, for the 41 non-included papers, a subset of 35 documents was selected for abstract screening based on their title analysis. The objective was to treat these papers and determine the level of information loss by not including some of them. After the examination, it was possible to verify that seven articles on the subjects of ethics [30], [32], [33], fairness [42] and transparency [22], [24], [28] provided relevant insights. In this case, these insights were considered and merged with those from the included papers to mitigate the loss of information, avoid bias and enrich this work's insights.

Second, reversely, throughout the analysis, it was marked to be reviewed if a paper was included but did not provide useful information due to its scope not relying on concerns about A.I. development and application within the public sector. At the end of this analysis, 11 articles on the subjects of explainability [54], interpretability [55]–[57], transparency [58], fairness [59]–[61], and trustworthiness [62]–[64].

The first case shows that seven (or about 8% of the selected) papers with relevant information were not included in the review initially. Nevertheless, a second round of analysis could capture and correct this distortion, adding the potential missing relevant insights to this work.

The second case shows that 11 (or about 25% of the included) had no relevant information about the topic in which they were fit for many reasons. No information was lost in this case, but a few lessons could be learned. For example, the word "transparency", which can be regarded as a concern in A.I. development and the public sector, in general, led to include a paper [58] that focused on transparency in a broad public sector context since the NLP analysis did not consider the context of the word in the whole text. Also, the paper [59], which focused on a reflexive, mixed-methods analysis of research contributions, shortcomings, and prospects throughout the four years of the FAccT Conference, was included as containing relevant information regarding fairness. Further, the papers [55]–[57] were included as having relevant information regarding interpretability but did not have, leading to the suppression of insights regarding this concern (the worst situation).

### 4.3 Results

This work proposed using an advanced search jointly with NLP and TM in two databases to support the selection of papers to be analysed. The results showed that the most frequent concern is fairness, and the most prominent topic is trustworthiness. Despite the abovementioned limitations, these results are consistent with the remaining literature and other resources not explored in this work.

First, fairness, which refers to correct algorithmic bias, has been firmly approached in some social sectors, such as justice. Also, it has been widely discussed in relevant conferences, movies, and documentaries, such as Coded Bias (2020). In addition, trustworthiness, which refers to avoiding unwanted side-effects generated by A.I. algorithms, has been highly explored through guidelines and reports from private, public and international organisations worldwide. Trustworthiness in the context of A.I. has been so significant that the European Commission has created the European Centre for Algorithmic Transparency (ECAT) to "contribute to a safer, more predictable and trusted online environment for people and business".

Second, besides these direct results, it was possible to gather relevant insights regarding data privacy, ethics, interpretability, explainability, trustworthiness, and fairness that are expected to serve as reflection points for the readers in different ways. Further, these insights are expected to be used as a starting point for other research, aggregating still more value to this work in the future.

Third, data privacy was captured as the least prominent topic. Nevertheless, data privacy can be viewed as the mother of all concerns, even when the world is not digital as today, and since then, an extensive concern with the issue of data privacy in a broader context worldwide – before all concerns regarding A.I. came to the surface. For instance, historically, the first legal text to be issued in the world on data privacy was the Law of the German State of Hesse in 1970. Years later, but not so recently, the Lisbon Treaty, signed on 13 December 2007 and in force since 1 December 2009, changed the paradigm of the European Union on data privacy. More recently, on 27 April 2016, the European Union's General Data Protection Regulation (GDPR) was published. With the growth of A.I. applications, data privacy has also become a concern in this context, especially about applications that use citizens' data. Thus, even if implicitly, the concern with data privacy is embedded in most articles.

Finally, a relevant collection of documents focused on the practical results of A.I. applications in the public sector is provided in this work. They are recent documents (last two years) and reflect the most relevant and current concerns regarding A.I. development, and, in many cases, they describe how these concerns have been overcome, which can be very useful as a benchmarking for those who are suffering from the same pain of developing and implementing A.I. applications within the public sector.

## 5 CONCLUSION

The methodology was efficient, saved time and captured relevant information. However, it needs to be improved, mainly in aggregating the context analysis to the search and increasing the number of databases such that it will be possible to enhance the results and generate more powerful insights.

Further, the design, development, deployment and use of A.I. systems require the participation of stakeholders from distinct domains such as technology, public, regulatory and users [65]. Further, principles such as those approached in this work are not individuals. They comprise ethical principles and are applied jointly to improve A.I. systems, ensuring human-centricity and public benefits with progressiveness and sustainability [37].

Finally, several frameworks have been proposed to assess data and A.I. models in this context. These frameworks have enabled organisations to modify their practices twofold: (i) regarding data, assessing them for training and testing in the models, besides post-implementation, and (ii) model-specifications characterisation, focusing particular attention on concerns such as accuracy, bias, consistency, transparency, interpretability, and fairness [66].

# A  APPENDICES

## A.1  Code 1 (search_files.py)

```python
import os
import fitz
import re
import pandas as pd
# Set the folder containing the PDF files - SHOULD BE CHANGED
folder_path = 'PATH/TO/FOLDER'
# Define the search expressions
expression_words = ['artificial intelligence', 'public sector']
# Define the search words
search_words = ['ethics', 'ethical', 'fair', 'fairness', 'data privacy',
'interpretable', 'interpretability', 'explainable', 'explainability',
'trust', 'trustworthy', 'trustworthness',' responsible A.I.',
'transparency']
# Initialise an empty list to store the results
results = []
# Loop through all files in the folder
for filename in os.listdir(folder_path):
    if filename.endswith('.pdf'):
        # Get the full path to the PDF file
        file_path = os.path.join(folder_path, filename)
        # Open the PDF file
        with fitz.open(file_path) as pdf_file:
            # Read the contents of the file
            text = ''
            for page in pdf_file:
                text += page.get_text().lower()
            # Check if the file contains the two expressions and at least one search word
            contains_expressions = all(expression in text for expression in expression_words)
            contains_search_word = any(re.search(word, text) for word in search_words)
            # If contains...
            if contains_expressions and contains_search_word:
                # Add just the filename to the results list by append method
                results.append(os.path.basename(file_path))
# Convert the results list to a DataFrame
df = pd.DataFrame(results, columns=['Filename'])
# Save the results to an Excel file - NAME SHOULD BE CHANGED
df.to_excel('NAME_OF_FILE.xlsx', index=False)
# Print a message indicating that the file was saved
print('Results saved to an Excel file.')
```

## A.2 Code 2 (count_words_ACM_searched_filtered.py)

```python
import os
import fitz
import pandas as pd
# List of words to search for
search_words = ['ethics', 'ethical', 'fair', 'fairness', 'data privacy',
'interpretable', 'interpretability', 'explainable', 'explainability',
'trust', 'trustworthy', 'trustworthiness', 'responsible A.I.',
'transparency']
# Function to search for occurrences of search_words in a single PDF file
def search_file(file_path):
    doc = fitz.open(file_path)
    occurrences = [0] * len(search_words)
    for page in doc:
        text = page.get_text()
        for i, word in enumerate(search_words):
            count = text.lower().count(word.lower())
            occurrences[i] += count
    doc.close()
    return occurrences
# Create a list to hold the results for each file
results = []
# Iterate over all PDF files in the specified directory - SHOULD BE CHANGED
directory = 'PATH/TO/FOLDER'
for file_name in os.listdir(directory):
    if file_name.endswith('.pdf'):
        file_path = os.path.join(directory, file_name)
        occurrences = search_file(file_path)
        results.append([file_name] + occurrences)
# Convert the results list to a pandas DataFrame
df = pd.DataFrame(results, columns=['File'] + search_words)
# Calculate the averages
averages = {}
for word in search_words:
    word_occurrences = df[word].values
    word_averages = word_occurrences[word_occurrences > 0].mean()
    averages[word] = round(word_averages, 0)
# Create a new DataFrame to hold the filtered results
filtered_results = []
for index, row in df.iterrows():
    file_name = row['File']
    for word in search_words:
        if row[word] >= averages[word]:
            filtered_results.append([file_name, word, row[word]])
```

```python
# Convert the filtered results list to a pandas DataFrame
filtered_df = pd.DataFrame(filtered_results, columns=['File', 'Word', 'Occurrences'])
# Save the filtered results to an Excel file – NAME SHOULD BE CHANGED
filtered_df.to_excel('NAME_OF_FILE.xlsx', index=False)
# Print a message indicating that the file was saved
print('Results saved to an Excel file.')
```

## A.3 Code 3 (count_words_ACM_searched_filtered.py)

```python
import os
import pandas as pd
import fitz
# Set the folder containing the PDF files - SHOULD BE CHANGED
folder_path = 'PATH/TO/FOLDER'
# Define the search words
search_words = ['ethics', 'ethical', 'fair', 'fairness', 'data privacy', 'interpretable', 'interpretability', 'explainable', 'explainability', 'trust', 'trustworthy', 'trustworthiness',' responsible A.I.', 'transparency']
# Create an empty dictionary to store the results
results = {}
# Loop through all files in the folder
for filename in os.listdir(folder_path):
    if filename.endswith('.pdf'):
        # Open the PDF file
        with fitz.open(os.path.join(folder_path, filename)) as doc:
            # Read the contents of the file
            text = ''
            for page in doc:
                text += page.get_text()
            # Search for occurrences of the search words in the text
            for word in search_words:
                count = text.lower().count(word.lower())
                # Add the count to the dictionary
                if word in results:
                    results[word] += count
                else:
                    results[word] = count
# Create a pandas DataFrame from the results dictionary
df = pd.DataFrame(list(results.items()), columns=['Word', 'Occurrences'])
# Save the DataFrame to an Excel file - NAME SHOULD BE CHANGED
df.to_excel('NAME_OF_FILE.xlsx', index=False)
# Print a message indicating that the file was saved
print('Results saved to an Excel file.')
```